\documentstyle[12pt,rotate]{article}

\input epsf.sty

\newcommand{\inclfig}[2]{\mbox{\epsfxsize=#1cm \epsfbox{#2.ps}}}
\newcommand{\insertfig}[2]{\mbox{\epsfxsize=#1cm \epsfbox{#2.eps}}}
\newcommand{\g}{{\sl g}}

\begin{document}
\begin{titlepage}

\centerline{\large \bf Two-loop effects in the evolution
                       of non-forward distributions.}

\vspace{10mm}

\centerline{\bf A.V. Belitsky$^{a,b,}$\footnote{Alexander von
            Humboldt Fellow.}, D. M\"uller$^{a,c}$,
            L. Niedermeier$^{a}$, A. Sch\"afer$^{a}$}

\vspace{18mm}

\centerline{\it ${^a}$Institut f\"ur Theoretische Physik, Universit\"at
                Regensburg}
\centerline{\it D-93040 Regensburg, Germany}
\centerline{\it ${^b}$Bogoliubov Laboratory of Theoretical Physics,
                Joint Institute for Nuclear Research}
\centerline{\it 141980, Dubna, Russia}
\centerline{\it ${^c}$Institute of Theoretical Physics, Leipzig University}
\centerline{\it 04109 Leipzig, Germany}

\vspace{20mm}

\centerline{\bf Abstract}

\hspace{0.5cm}

We study the effects of next-to-leading order corrections on the evolution
of the twist-two non-forward parton distribution functions in the flavour
non-singlet sector. It is found that the deviation from leading order
evolution is small for all values of the parton momentum fraction variable
for moderately large values of the scale parameter.

\vspace{5cm}


\vspace{0.5cm}


\end{titlepage}

%%%%%%%%%%%%%%%%%%%%%%%%%%%%%%%%%%%%%%%%%%%%%%%%%%%%%%%%%%%%%%%%%%%%%
%\section{Introduction.}
%%%%%%%%%%%%%%%%%%%%%%%%%%%%%%%%%%%%%%%%%%%%%%%%%%%%%%%%%%%%%%%%%%%%%

\noindent {\it 1. Introduction.} Recently there was significant
progress in the perturbative QCD approach to deeply virtual
Compton scattering --- a process which allows to access the
so-called non-forward parton distribution functions
\cite{Ji96,Rad96,Rad97,GeyDitHorMueRob88}. The latter
possess hybrid properties: in different regions of phase
space they share features with ordinary parton densities
of deep inelastic scattering and with exclusive distribution
amplitudes. They smoothly interpolate between these two
limits. At the present stage all the required perturbative
inputs are available, i.e.\ one-loop coefficient
functions \cite{JiOs97,BelMue97a,Mank97,BelSch98b,JiOs98} and
two-loop anomalous dimensions of the moments of the non-forward
functions \cite{BelMue98c,BelMue98d}, which make possible its
analysis in next-to-leading order (NLO) of perturbation theory
in the flavour singlet channel. In the above list of corrections
the latter were derived within a formalism which allows to
determine the eigenfunctions of the two-loop generalized
Efremov-Radyushkin-Brodsky-Lepage (ER-BL) evolution equations in
closed analytical form. This is a crucial result which is
not accessible in the direct calculation of NLO kernels.

The leading order solution of the evolution equations, which
are governed by the kernels evaluated in Refs.\
\cite{Ji96,Rad96,Rad97,FFGS97,Blum97,BalRad97,BelMue97a,HooJi98},
in terms of the conformal partial wave expansion was given in
Refs.\ \cite{Rad97,BelMue97a,BelGeyMueSch97e} (see also
\cite{ManPilWei97} and for a direct numerical integration of
evolution equation Ref.\ \cite{FFGS97}). In two-loop approximation
the corrections to the eigenfunctions were derived in
\cite{BelMue97a,BelMue98d}. In the present investigation we will
use these results for an explicit study of the magnitude of these
effects on the evolution of the flavour non-singlet non-forward
parton distribution function versus the LO evolution considered
by us previously \cite{BelGeyMueSch97e}.

%%%%%%%%%%%%%%%%%%%%%%%%%%%%%%%%%%%%%%%%%%%%%%%%%%%%%%%%%%%%%%%%%%%%%
%\section{Solution of the two-loop evolution equation.}
%%%%%%%%%%%%%%%%%%%%%%%%%%%%%%%%%%%%%%%%%%%%%%%%%%%%%%%%%%%%%%%%%%%%%

\noindent {\it 2. Solution of the two-loop evolution equation.} To
start with let us first describe the formalism and spell out our
conventions. Here we adopt the definition of the non-forward parton
distributions introduced by Radyushkin \cite{Rad96,Rad97} which
fulfill the evolution equation
\begin{equation}
\label{EvolEq}
\mu^2\frac{d}{d\mu^2} {\cal O} (x, \zeta)
= \int d x ' K
\left( x , x ', \zeta \left| \alpha_s (Q^2) \right)\right.
{\cal O} (x ', \zeta) ,
\end{equation}
where the kernel $K \left( x , x ', \zeta \left| \alpha_s (Q^2)
\right)\right. $ can be calculated order by order in perturbation
theory. Since the leading order evolution
equation can be diagonalized with the help of the conformal
operators it is convenient to employ this partial conformal
wave expansion also beyond leading order although the Gegenbauer
polynomials are not the eigenfunctions of the two-loop generalized
ER-BL equation. Namely, the solution of the Eq. (\ref{EvolEq}) can
be written in the form
\begin{equation}
\label{ConfExp}
{\cal O} (x, \zeta, Q^2)
= \sum_{j = 0}^{\infty}
\phi_j \left( x, \zeta \left| \alpha_s (Q^2) \right)\right.
\widetilde{\cal O}_j ( \zeta, Q^2 ),
\end{equation}
where the multiplicative renormalizable moments evolve as
follows
\begin{equation}
\label{Solution}
\widetilde{\cal O}_j ( \zeta, Q^2 )
= \exp
\left\{
- \frac{1}{2} \int_{Q_0^2}^{Q^2} \frac{d \tau}{\tau}
\gamma_j^{\rm D} \left( \alpha_s (\tau) \right)
\right\}
\widetilde{\cal O}_j ( \zeta, Q_0^2 ) ,
\end{equation}
with the forward anomalous dimensions we need at ${\cal O} (\alpha_s^2)$
accuracy $\gamma_j^{\rm D} \left( \alpha_s \right)
= \frac{\alpha_s}{2 \pi} \ \gamma_j^{(0)}
+ \left( \frac{\alpha_s}{2 \pi} \right)^2 \gamma_j^{(1)} + \dots$,
where $\gamma_j^{(0)} = - C_F \left( 3 +
\frac{2}{( j + 1 )( j + 2 )} - 4 \psi( j + 2 ) + 4 \psi(1) \right)$
and $\gamma_j^{(1)}$ can be found in Ref.\ \cite{LopYnd79}. We have
chosen the initial condition so that there are no radiative
corrections at the low normalization point $Q^2_0$ so that the
$\widetilde{\cal O}_j (\zeta, Q_0^2)$ are given by ordinary
Gegenbauer moments of the non-forward distribution which are
related to the matrix elements of the tree level conformal
operators by
\begin{equation}
\label{MatrixEl}
\widetilde{\cal O}_j (\zeta, Q_0^2)
= \int dx \ C_j^{3/2} \left( 2 \frac{x}{\zeta} - 1 \right)
{\cal O} (x, \zeta, Q_0^2)
= \frac{1}{\zeta^j}
\langle h'|
\left.
\bar{\psi} (i \partial_+)^j {\mit\Gamma}
C^{3/2}_j \left( {\stackrel{\leftrightarrow}{D}_+}
/ {\partial_+} \right) \psi
\right|_{Q_0^2}
| h \rangle .
\end{equation}
The problem is reduced to finding the correction to the
eigenfunction. It was solved in our previous studies
\cite{BelMue97a,BelMue98c,BelMue98d,Mue94}
\begin{equation}
\label{CorrConfWave}
\phi_j \left( x ,\zeta \left| \alpha_s (Q^2) \right)\right.
= \phi_j ( x, \zeta )
+ \frac{\alpha_s (Q^2)}{2 \pi} \sum_{k = j + 2}^{\infty}
\phi_k ( x, \zeta ) {\mit\Phi}_{kj} \left( \alpha_s (Q^2) \right) ,
\end{equation}
where the LO partial conformal waves are defined via the Gegenbauer
polynomials
\begin{equation}
\label{ConfWave}
\phi_j (x, \zeta)
\equiv
\frac{1}{ N_j }
\frac{x}{\zeta^2} \left( 1 - \frac{x}{\zeta} \right)
C_j^{3/2} \left( 2 \frac{x}{\zeta} - 1 \right),
\quad\mbox{with}\quad
N_j = \frac{(j + 1)(j + 2)}{4 (2 j  + 3)} .
\end{equation}
The function ${\mit\Phi}_{jk}$ is \cite{BelMue97a,BelMue98c,BelMue98d,Mue94}
\begin{equation}
{\mit\Phi}_{jk} \left( \alpha_s (Q^2) \right)
= S_{jk} \left( \alpha_s (Q^2) \right)
\left\{
d_{jk} \left( \gamma_k^{(0)} - \beta_0 \right)
- g_{jk}
\right\},
\end{equation}
with
\begin{eqnarray}
d_{jk}
&=& - \frac{1}{2}[ 1 + ( - 1)^{j - k} ]
\frac{(2k + 3)}{(j - k)(j + k + 3)},\\
g_{jk}
&=& 2 C_F \ d_{jk}
\left\{
2 A_{jk}
+ ( A_{jk} - \psi (j + 2) + \psi (1) )
\frac{(j - k)(j + k + 3)}{(k + 1)(k + 2)}
\right\}, \\
\mbox{with}&&\!\!\!\! A_{jk} = \psi \left( \frac{j + k + 4}{2} \right)
- \psi \left( \frac{j - k}{2} \right)
+ 2 \psi ( j - k ) - \psi ( j + 2 ) - \psi (1) . \nonumber
\end{eqnarray}
The factor $S_{jk}$ appears as result of the evolution of the
coupling constant and reads
\begin{equation}
S_{jk} \left( \alpha_s (Q^2) \right)
= \frac{ \gamma_j^{(0)} - \gamma_k^{(0)} }{
\gamma_j^{(0)} - \gamma_k^{(0)} + \beta_0 }
\left(
1
- \left( \frac{\alpha_s(Q_0^2)}{\alpha_s (Q^2)} \right)^{
1 + \left( \gamma_j^{(0)} - \gamma_k^{(0)} \right)/\beta_0}
\right) ,
\end{equation}
with $\beta_0 = \frac{4}{3} T_F N_f - \frac{11}{3}C_A$.

Obviously, taken as they stand the above Eqs.\
(\ref{ConfExp},\ref{CorrConfWave},\ref{ConfWave}) are valid only
for the non-forward distributions with support $0 \leq x \leq \zeta$
since the Gegenbauer polynomials $C_j^\nu (2 x - 1)$ form a complete set
only on the interval $x \in [0,1]$. But as has been noted in Ref.
\cite{Rad97} that means that we can understand the above expansion
only in a restricted sense. Namely, we have to represented it in the
form \cite{BelSch98b}
\begin{equation}
\label{MathDistr}
\left[
\frac{x}{\zeta^2}
\left( 1 - \frac{x}{\zeta} \right)
\right]^{\nu - 1/2}
C^\nu_j \left( 2 \frac{x}{\zeta} - 1 \right)
= 2^{1 - 2\nu}
\frac{\Gamma \left( \frac{1}{2} \right) \Gamma (j + 2 \nu)}{
\Gamma (\nu) \Gamma (j + \nu + \frac{1}{2}) \Gamma (j + 1)}
\int_{0}^{1} dt (t \bar t)^{j + \nu - 1/2}
\delta^{(j)} (\zeta t - x),
\end{equation}
and treat the RHS as a mathematical distribution\footnote{For
this reason we omit the spectral constraint which appears on the LHS
as a result of integration.}. In order to circumvent this disadvantage
we expand the non-forward distribution in a series of polynomials
${\cal P}_j (x)$ orthogonal in the region $0 \leq x \leq 1$. Then
the expansion coefficients will be given as convolution of
${\cal P}_j (x)$ with the RHS of Eq.\ (\ref{MathDistr}).

Note that we can exploit any set of orthogonal polynomials
${\cal P}_j (x)$ available. Namely, the general expansion looks
like
\begin{equation}
\label{PolExp}
{\cal O} (x, \zeta, Q^2) = \sum_{j = 0}^{\infty}
\widetilde{\cal P}_j (x) {\cal M}_j^{\cal P} (\zeta, Q^2),
\end{equation}
where the conjugated polynomials $\widetilde{\cal P}_j (x)$ are
defined such that
\begin{equation}
\int_{0}^{1} dx\ \widetilde{\cal P}_j (x) {\cal P}_k (x)
= \delta_{jk}.
\end{equation}
And the moments ${\cal M}_j^{\cal P} (\zeta, Q^2)$ are given by a
finite sum
\begin{equation}
{\cal M}_j^{\cal P} (\zeta, Q^2)
= \sum_{k = 0}^{j} E_{jk}^{\cal P} (\zeta)
{\cal O}_k (\zeta, Q^2),
\end{equation}
with ${\cal O}_j$-moments expressed in terms of the original ones as
follows
\begin{equation}
{\cal O}_j (\zeta, Q^2)
= \widetilde{\cal O}_j (\zeta, Q^2)
+ \frac{\alpha_s (Q^2)}{2 \pi}
\sum_{k = 0}^{j - 2} {\mit\Phi}_{jk} \left( \alpha_s (Q^2) \right)
\widetilde{\cal O}_k (\zeta, Q^2).
\end{equation}
Taking into account the above observation, the expansion coefficients
are given by the integral
\begin{equation}
E_{jk}^{\cal P} (\zeta) = \int_{0}^{1} dx\
\frac{x \bar x}{N_k} C_k^{3/2} ( 2x - 1 ) {\cal P}_j ( \zeta x ).
\end{equation}
Let us repeat that we can use any appropriate polynomials for
this purpose. The criterion for choosing a specific one is thus
the fastest convergence of the series. Below we give the result
for the expansion coefficients of the Jacobi polynomials. Namely,
\begin{eqnarray}
&&{\cal P}_j (x) = P_j^{(\alpha, \beta)} (2x - 1),
\quad
\widetilde{\cal P}_j (x)
= \frac{\bar x^\alpha x^\beta}{n_j (\alpha, \beta)}
P_j^{(\alpha, \beta)} (2x - 1), \nonumber\\
&&\mbox{with}\quad
n_j (\alpha, \beta)
= \frac{\Gamma (j + \alpha + 1) \Gamma (j + \beta + 1)}{
(2j + \alpha + \beta + 1) j! \Gamma (j + \alpha + \beta + 1)}
\end{eqnarray}
The expansion coefficients can easily be obtained by the methods
developed in Ref. \cite{BelMue98d} and read
\begin{eqnarray}
E_{jk}^J (\zeta) \!\!\!&=&\!\!\! (- 1)^{j - k} \theta_{jk}
\frac{\Gamma (k + 2)}{\Gamma (2k + 3)}
\frac{\Gamma (j + \beta + 1)}{\Gamma (k + \beta + 1)}
\frac{\Gamma (j + k + \alpha + \beta + 1)}{\Gamma (j - k)
\Gamma (j + \alpha + \beta + 1 )} \nonumber\\
&&\qquad\qquad\times 2 \zeta^k {_3 F_2}
\left.\left(
{ -j + k , j + k + \alpha + \beta + 1 , k + 2
\atop
2k + 4 , k + \beta + 1 }
\right| \zeta \right) .
\end{eqnarray}
The results for all other classic orthogonal polynomials immediately
follow from this expression. For special values of the parameters
the Jacobi polynomials coincide \cite{AbrSte65} either with
Gegenbauer\footnote{We have use them in our preliminary study of LO
evolution effects \protect\cite{BelGeyMueSch97e}.},
$P_j^{( \lambda - \frac{1}{2}, \lambda - \frac{1}{2})} (x)
= \frac{\Gamma (2 \lambda) \Gamma (j + \lambda + \frac{1}{2})}{
\Gamma (j + 2 \lambda) \Gamma (\lambda + \frac{1}{2})} C_j^\lambda (x)$,
or Legendre\footnote{This possibility has been discussed in Ref.\
\cite{ManPilWei97} but the authors did not manage to find an explicit
analytical expression for the expansion coefficients.}, $P_j^{(0,0)}
(x) = P_j (x)$, or Chebyshev polynomials of the first,
$P_j^{(- \frac{1}{2}, - \frac{1}{2})} (x) =
\frac{\Gamma (j + \frac{1}{2})}{\sqrt{\pi} j!} T_j (x)$ and second
kind, $P_j^{(\frac{1}{2}, \frac{1}{2})} (x) =
\frac{2 \Gamma (j + \frac{3}{2})}{\sqrt{\pi} (j + 1)!} U_j (x)$.

The solution of the renormalization group equation (\ref{Solution})
in two-loop approximation can be written in the form
\begin{eqnarray}
\widetilde{\cal O}_j (\zeta, Q^2)
&=& \widetilde{\cal O}_j (\zeta, Q^2_0)
\left(
\frac{\alpha_s (Q_0^2)}{\alpha_s (Q^2)}
\right)^{ \gamma_j^{(0)} / \beta_0 }
\left(
\frac{ \beta_0 + \beta_1 \frac{\alpha_s (Q^2)}{4 \pi} }{
\beta_0 + \beta_1 \frac{\alpha_s (Q^2_0)}{4 \pi} }
\right)^{\left( \gamma_j^{(0)} / \beta_0
- 2 \gamma_j^{(1)} / \beta_1 \right)} \nonumber\\
&=& \widetilde{\cal O}_j (\zeta, Q^2_0)
\left(
\frac{\alpha_s (Q_0^2)}{\alpha_s (Q^2)}
\right)^{ \gamma_j^{(0)} / \beta_0 }
\left\{
1
+ \left( \frac{\beta_1 \, \gamma_j^{(0)} }{ 2 \beta_0^2}
- \frac{\gamma_j^{(1)}}{\beta_0} \right)
\frac{\alpha_s(Q^2) - \alpha_s(Q_0^2) }{2 \pi}
\right\} ,
\end{eqnarray}
where the expansion in the second line is done in order to treat
two-loop corrections to the evolution on the same footing as
one-loop corrections to the Wilson coefficients, which when both
are summed in the amplitude allows to minimize the renormalization
scheme dependence (see, for instance, the first paper in Ref.\
\cite{LopYnd79} and \cite{GluRey82}). The coupling constant in NLO
of perturbation theory can be approximated by
\begin{equation}
\alpha_s(Q^2) = - \frac{4\pi}{\beta_0
\ln ( Q^2 / \Lambda^2_{\overline{\rm MS}})}
\left(
1 + \frac{\beta_1}{\beta_0^2}
\frac{\ln \, \ln ( Q^2 / \Lambda^2_{\overline{\rm MS}} )}{
\ln ( Q^2 / \Lambda^2_{\overline{\rm MS}}) }
\right),
\end{equation}
where $\beta_1$ is the second coefficient in the expansion of the QCD
$\beta$-function $\frac{\beta}{\g} = \frac{\alpha_s}{4\pi}\ \beta_0
+ \left( \frac{\alpha_s}{4\pi} \right)^2 \beta_1 + \dots$ and it reads
$\beta_1 = \frac{10}{3} C_A N_f + 2 C_F N_f - \frac{34}{3} C_A^2$.

%%%%%%%%%%%%%%%%%%%%%%%%%%%%%%%%%%%%%%%%%%%%%%%%%%%%%%%%%%%%%%%%%%%%%
%\section{NLO evolution of the model distributions.}
%%%%%%%%%%%%%%%%%%%%%%%%%%%%%%%%%%%%%%%%%%%%%%%%%%%%%%%%%%%%%%%%%%%%%

\noindent {\it 3. NLO evolution of the model distributions.} In this
section we will use the results given above for explicit numerical
studies of the evolution of the non-forward parton distributions. To this
end we need an initial condition for the evolution equation. The most
adequate model for the low scale input functions was given by
Radyushkin in Ref.\ \cite{Rad98}. Namely, ${\cal O} (x, \zeta)$ is
defined in terms of the double distribution function $F(y, z)$
\cite{Rad97} via the following relation
\begin{equation}
\label{nontodouble}
{\cal O} (x, \zeta, Q^2_0)
=  \int_0^1 d y \int_0^1 d z \, F (y, z, Q^2_0) \,
\theta (1 - y - z) \delta (x - y - \zeta z).
\end{equation}
Here we have omitted the dependence on the $t$-channel momentum
transferred squared. Although for a massive target the formal
limit $t \to 0$ is not accessible for nonvanishing $\zeta$ due to
the kinematical restriction, $m_h^2 \zeta^2/\bar\zeta \leq - t$
\cite{AbrFraStr94,Rad97}, this condition does not affect the
results of the evolution.

\begin{figure}[t]
\unitlength1mm
\begin{center}
\begin{picture}(150,110)(0,0)
\put(0,60){\insertfig{6}{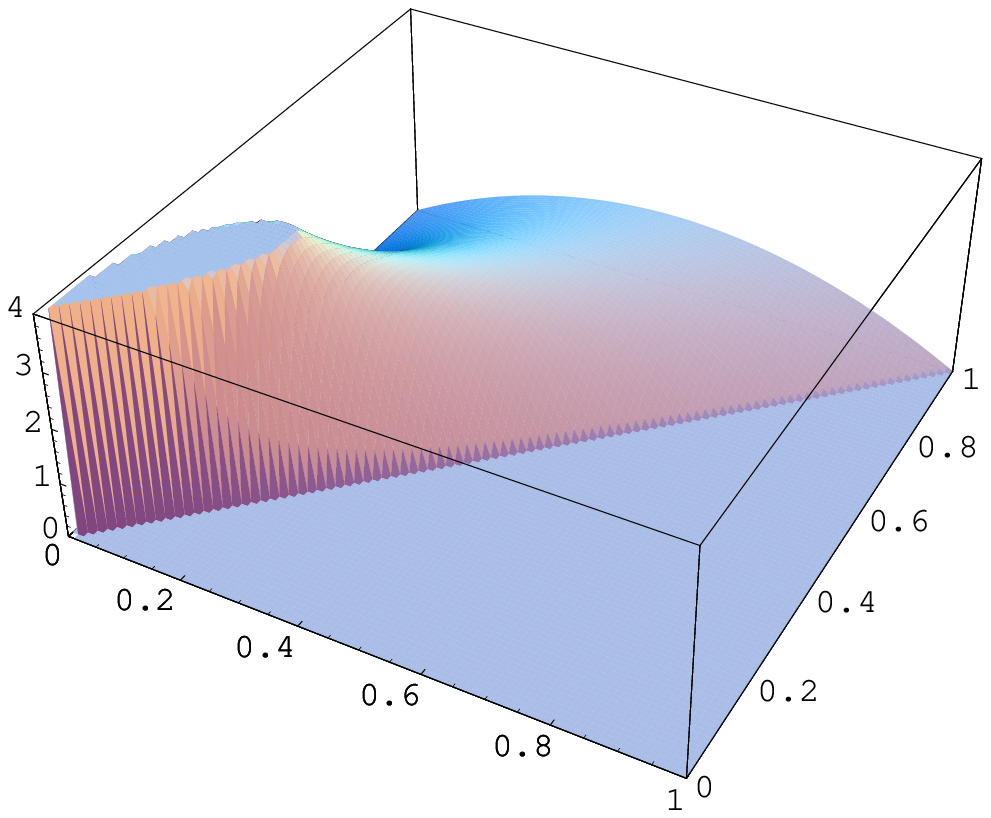}}
\put(50,97){(a)}
\put(20,60){$x$}
\put(55,70){$\zeta$}
\put(-10,73){
\rotate{$\scriptstyle {\cal O}^{{\rm NS}}\left(x,\zeta,Q^2\right)$}}
\put(83,58){\inclfig{5.3}{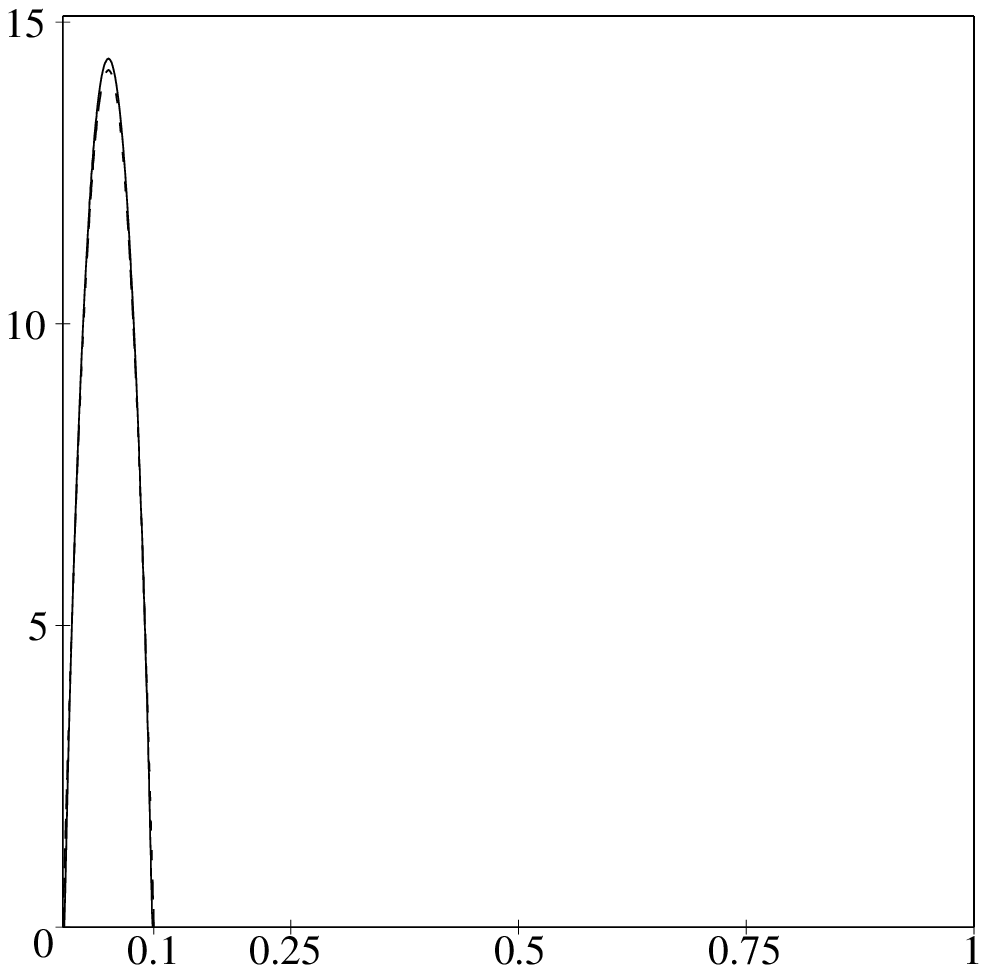}}
\put(126,102){(b)}
\put(110,55){$x$}
\put(75,73){
\rotate{$\scriptstyle {\cal O}^{{\rm NS}}\left(x,\zeta,Q^2\right)$}}
\put(-2,-2){\inclfig{5.3}{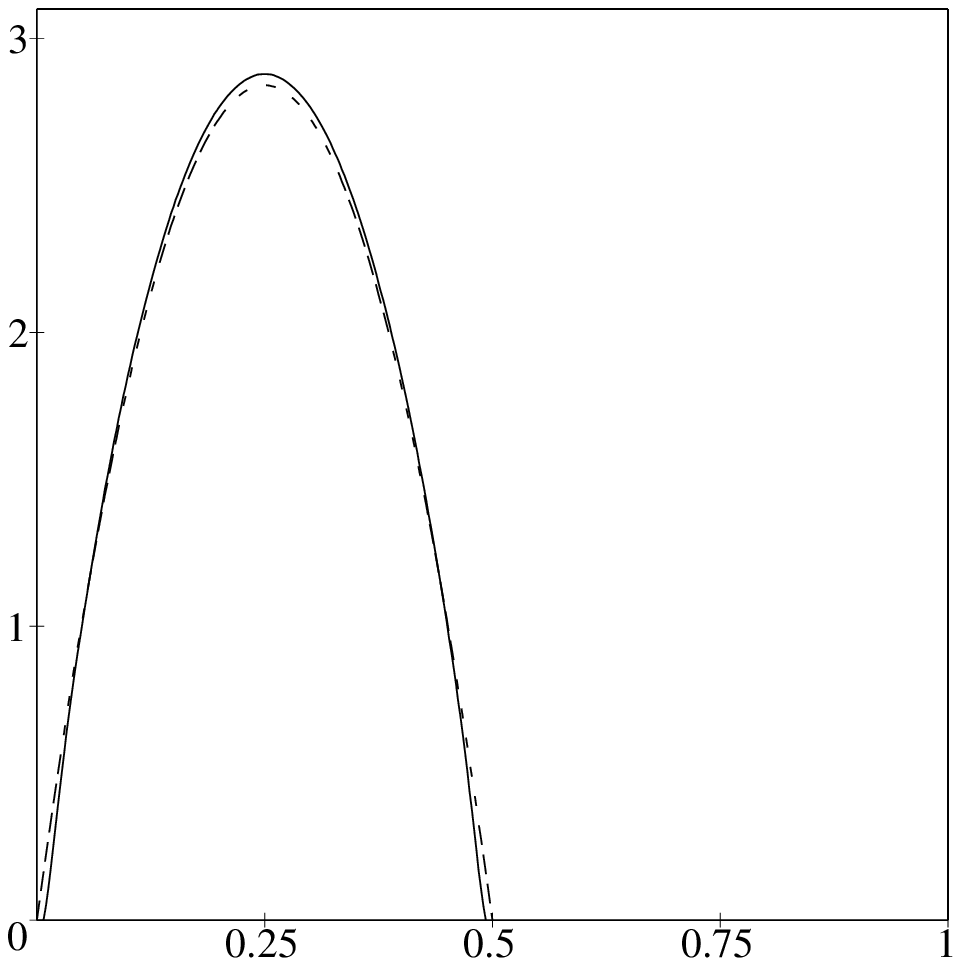}}
\put(41,43){(c)}
\put(25,-5){$x$}
\put(-10,13){
\rotate{$\scriptstyle {\cal O}^{{\rm NS}}\left(x,\zeta,Q^2\right)$}}
\put(83,-2){\inclfig{5.3}{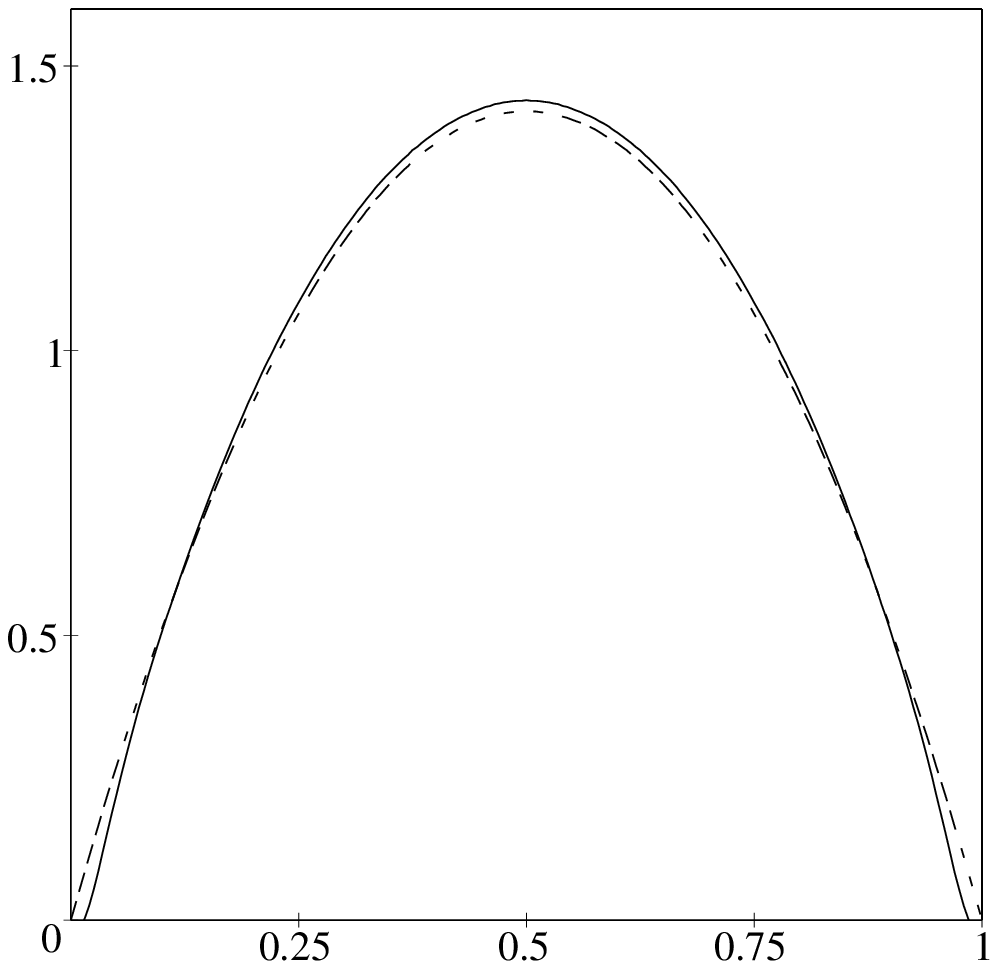}}
\put(126,43){(d)}
\put(110,-5){$x$}
\put(75,13){
\rotate{$\scriptstyle {\cal O}^{{\rm NS}}\left(x,\zeta,Q^2\right)$}}
\end{picture}
\end{center}
\caption{\label{Asydistribution} Evolution of the asymptotic
distribution function. The $x-\zeta$-dependence given by Eq.\
(\protect\ref{AsDis}) is shown in (a). The input distribution
(dashed-dotted line) was evolved in NLO approximation (solid line) up
to $Q^2=100\ {\rm GeV}^2$ for the skewdness parameters $\zeta = 0.1$
(b), $\zeta = 0.5$ (c) and $\zeta = 1.0$ (d).}
\end{figure}

For our study we accept the following model for the
double distribution function corresponding to the non-singlet
function ${\cal O} (x, \zeta, Q^2_0)$ \cite{Rad98}
\begin{equation}
\label{DDfunction}
F (y, z, Q^2_0) = q (y, Q^2_0) \pi (y, z),
\end{equation}
with the plausible profile functions $\pi (y, z)$
\begin{equation}
\label{realprofile}
\pi (y, z) = 6 \frac{z}{\bar y^3} ( \bar y - z ) .
\end{equation}
In Eq. (\ref{DDfunction}) the function $q (y, Q^2_0)$ is an ordinary
forward parton density measured in deep inelastic scattering taken at a
low normalization point. We will also consider the asymptotic distribution
functions which although phenomenologically probably irrelevant serves
as a good probe for the net evolution effects in NLO approximation since
its scale dependence is governed by the non-diagonal elements of the
anomalous dimension matrix of the conformal operators which appear only
beyond leading order. It is given by the first term in the expansion
(\ref{ConfExp}) \cite{Rad98,BelGeyMueSch97e}
\begin{equation}
\label{AsDis}
{\cal O}_{\rm as} (x, \zeta, Q^2_0)
= 6 \frac{x}{\zeta^2} \left( 1 - \frac{x}{\zeta} \right)
\theta ( \zeta - x ) {\cal O}_0 (\zeta, Q^2_0),
\end{equation}
(see Fig.\ \ref{Asydistribution} (a)) where ${\cal O}_0
(\zeta, Q^2_0)$ (see Eq. (\ref{MatrixEl})) is a matrix element of
the conserved local (axial-)vector current, ${\cal O}_0 (\zeta, Q^2_0)
= \langle h'| \bar \psi (0) {\mit\Gamma} \psi (0) |h \rangle$ with
${\mit\Gamma} = \gamma_+, \gamma_+ \gamma_5$. Therefore, it depends
neither on the skewedness parameter nor on the renormalization point
$Q^2_0$.

\begin{figure}[t]
\unitlength1mm
\begin{center}
\begin{picture}(150,110)(0,0)
\put(0,60){\insertfig{6}{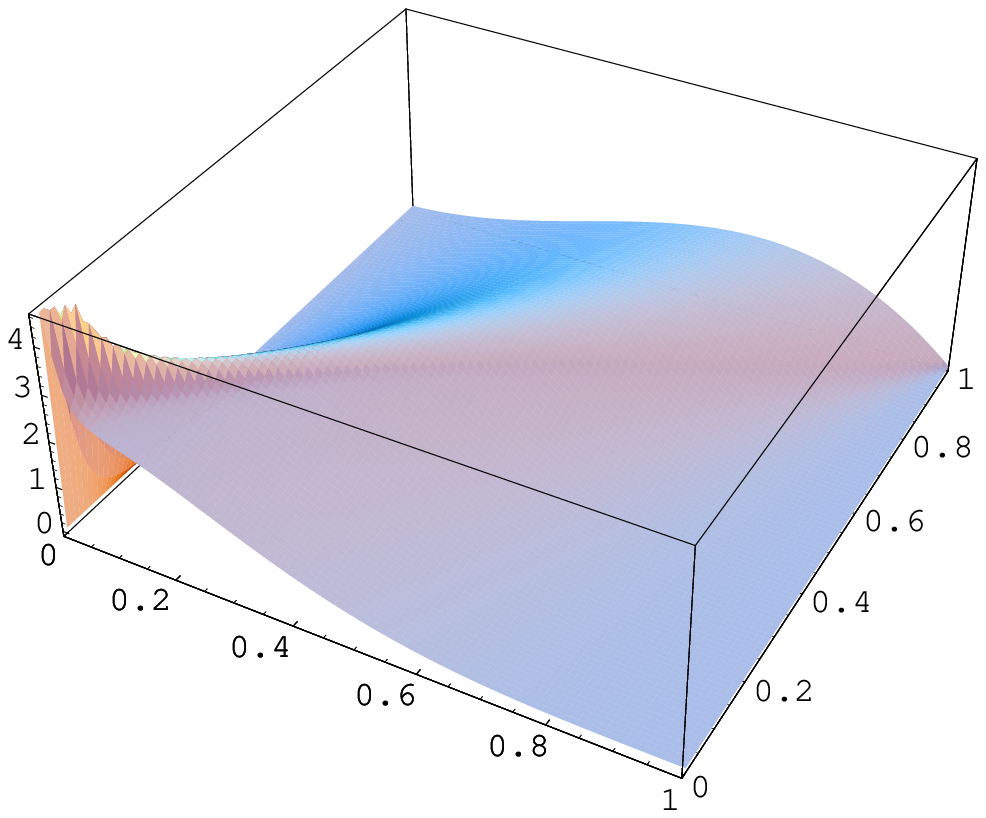}}
\put(50,97){(a)}
\put(20,60){$x$}
\put(55,70){$\zeta$}
\put(-10,73){
\rotate{$\scriptstyle {\cal O}^{{\rm NS}}\left(x,\zeta,Q^2\right)$}}
\put(83,58){\inclfig{5.3}{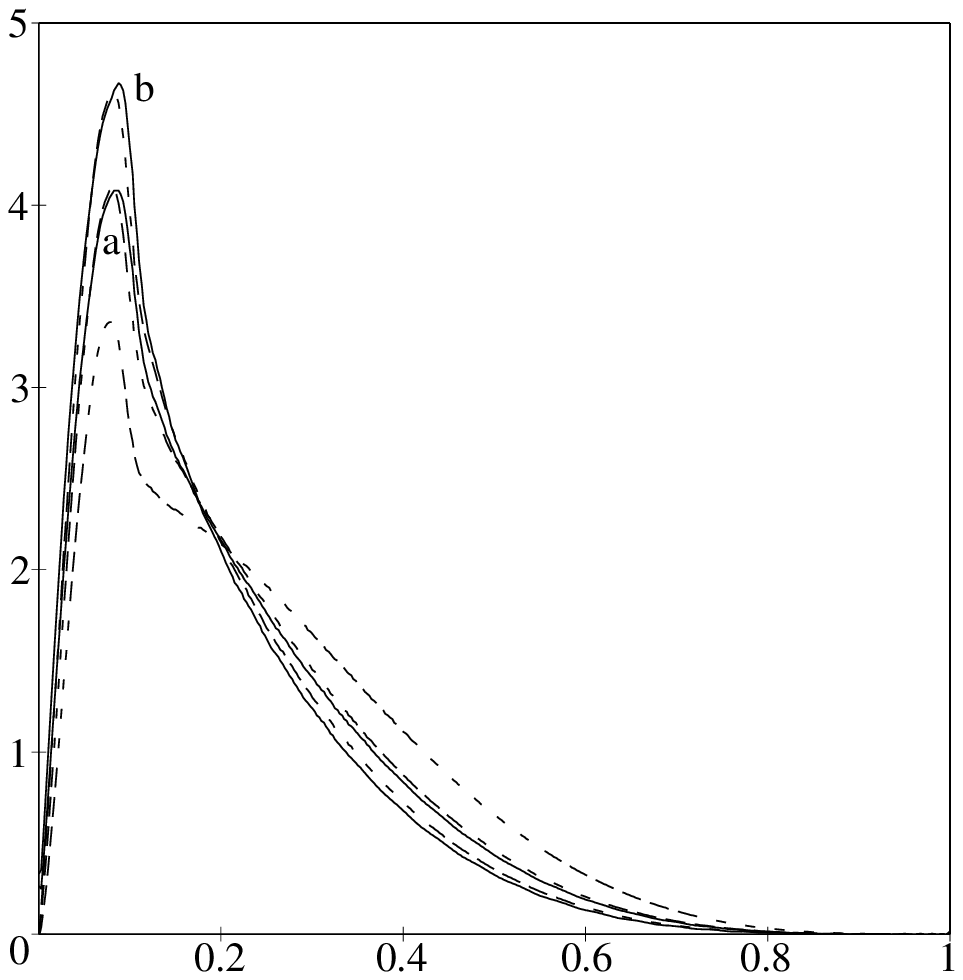}}
\put(126,102){(b)}
\put(110,55){$x$}
\put(75,73){
\rotate{$\scriptstyle {\cal O}^{{\rm NS}}\left(x,\zeta,Q^2\right)$}}
\put(-2,-2){\inclfig{5.3}{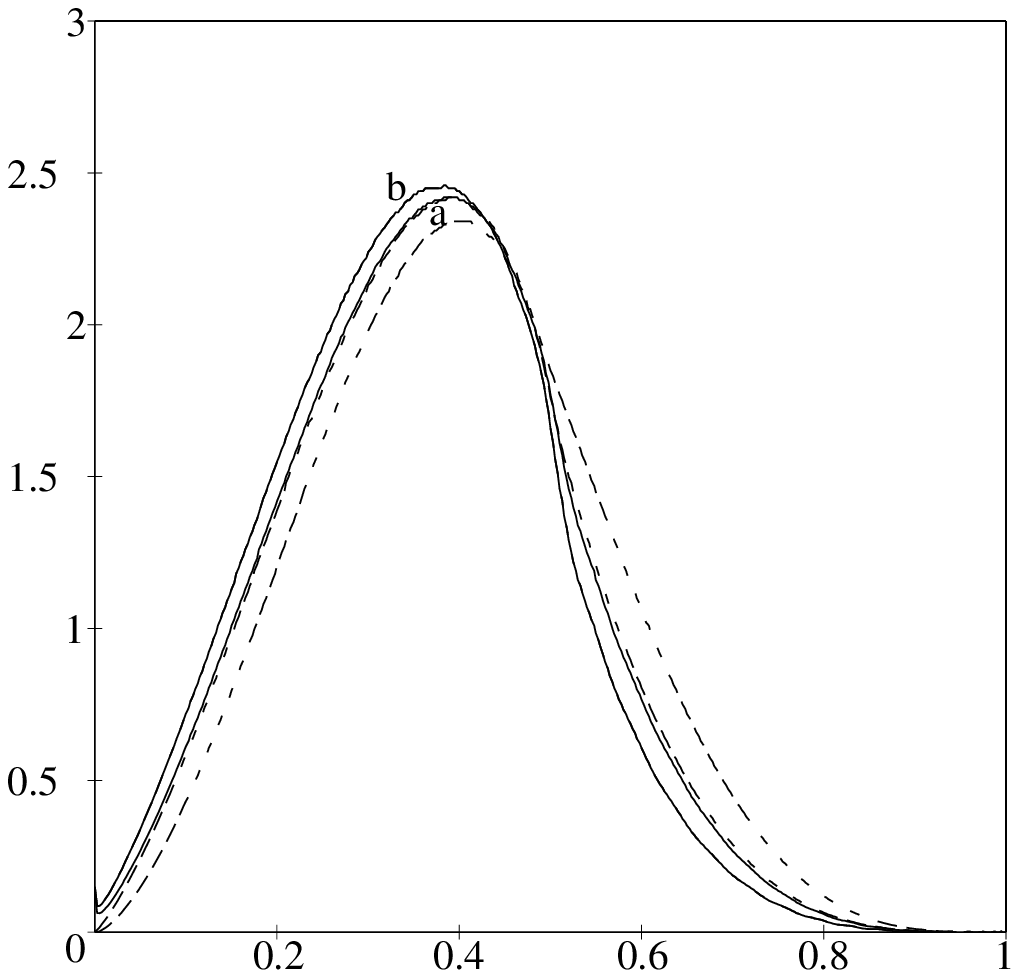}}
\put(41,43){(c)}
\put(25,-5){$x$}
\put(-10,13){
\rotate{$\scriptstyle {\cal O}^{{\rm NS}}\left(x,\zeta,Q^2\right)$}}
\put(83,-2){\inclfig{5.3}{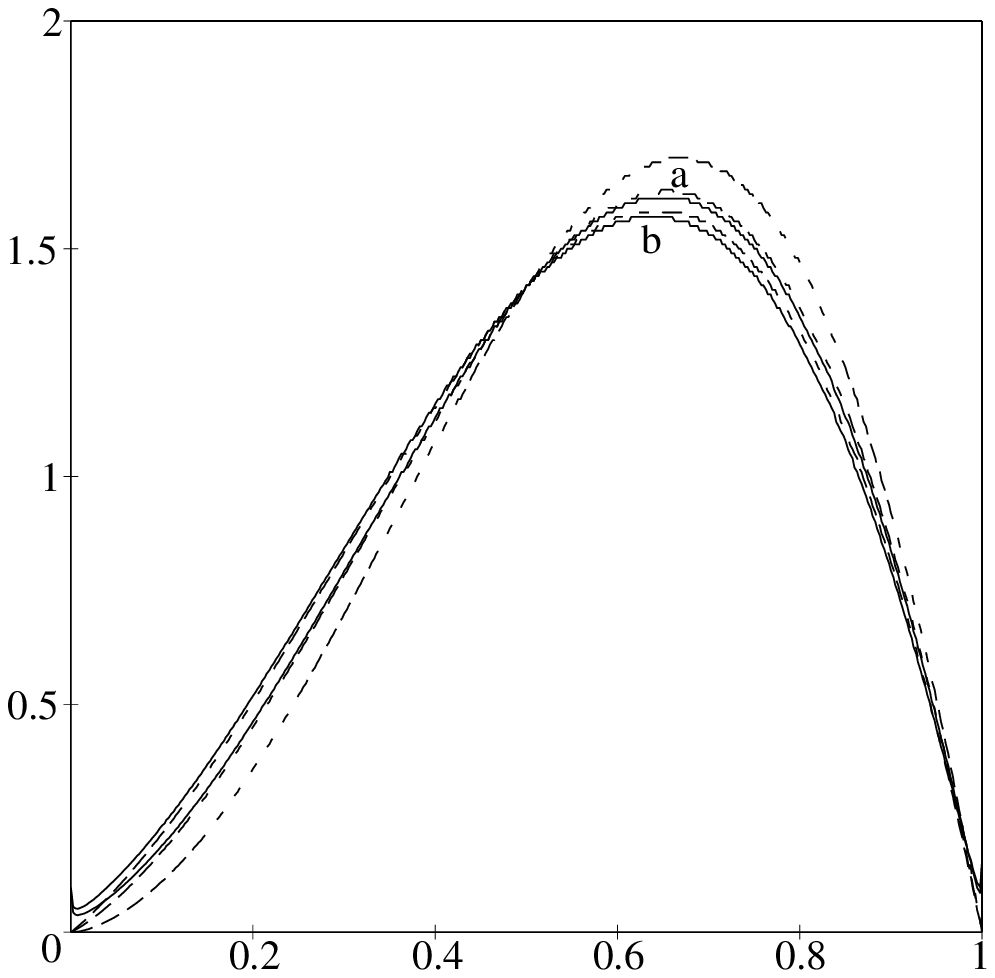}}
\put(126,43){(d)}
\put(110,-5){$x$}
\put(75,13){
\rotate{$\scriptstyle {\cal O}^{{\rm NS}}\left(x,\zeta,Q^2\right)$}}
\end{picture}
\end{center}
\caption{\label{RadyushkinModel} 3D shape of the Radyushkin's model (a).
Evolution of input distribution (dashed-dotted line) in leading (dashed
lines) and next-to-leading order (solid lines) with $\zeta =0.1$ (b),
$\zeta = 0.5$ (c), $\zeta = 1.0$ (d) and $Q^2=10\ {\rm GeV}^2 $ (curve a),
$Q^2=100\ {\rm GeV}^2 $ (curve b).}
\end{figure}

Since in present paper we are studying only the non-singlet evolution
we consider the combination
\begin{equation}
q^{\rm NS} (x) = u (x) - d (x),
\end{equation}
and take as initial condition the following CTEQ4M parametrizations
for the $u$ and $d$-quark distributions \cite{Lai97}
\begin{eqnarray}
u (x) &=& 1.344 x^{-0.499} (1 - x)^{3.689} (1 + 6.042 x^{0.873}),\\
d (x) &=& 0.640 x^{-0.499} (1 - x)^{4.247} (1 + 2.690 x^{0.333}) ,
\end{eqnarray}
at an input scale $Q_0 = 1.6\ {\rm GeV}$. Then the above model reads
form these and is given in Fig.\ \ref{RadyushkinModel} (a). Note that
the factor ${\cal O}_0 (\zeta, Q^2_0)$ in the asymptotic distribution
is defined by the first moments of $q^{\rm NS} (x)$, ${\cal O}_0
(\zeta, Q^2_0) = \int_{0}^{1} dx\ q^{\rm NS} (x)$. Let us mention
that the models we use satisfy the positivity constraints\footnote{Note
an extra factor of $1 / \sqrt{\bar\zeta}$ in Eq.\ (\protect\ref{positiv})
found in Ref.\ \cite{Rad98} which was missed in \cite{PirSofTer98}.}
derived in Ref.\ \cite{PirSofTer98,Rad98}
\begin{equation}
\label{positiv}
{\cal O}(x, \zeta)
\leq
\sqrt{ q (x) q \left( ( x - \zeta )/\bar \zeta \right) / \bar\zeta} .
\end{equation}
Moreover, it is useful to mention that Eq.\ (\ref{nontodouble}) with
profile defined by (\ref{realprofile}) saturates the constraint
inequality (\ref{positiv}) in the region of its validity $x > \zeta$.

%%%%%%%%%%%%%%%%%%%%%%%%%%%%%%%%%%%%%%%%%%%%%%%%%%%%%%%%%%%%%%%%%%%%%
%                         Figure 1
%%%%%%%%%%%%%%%%%%%%%%%%%%%%%%%%%%%%%%%%%%%%%%%%%%%%%%%%%%%%%%%%%%%%%

%%%%%%%%%%%%%%%%%%%%%%%%%%%%%%%%%%%%%%%%%%%%%%%%%%%%%%%%%%%%%%%%%%%%%
%                         Figure 2
%%%%%%%%%%%%%%%%%%%%%%%%%%%%%%%%%%%%%%%%%%%%%%%%%%%%%%%%%%%%%%%%%%%%%

Now we are in a position to present the numerical results for the
evolution of the models introduced above. The appropriate values
for the parameters which we did not mention so far are $N_f=4 $,
$\Lambda_{\overline{\rm MS}} = 220\ {\rm MeV}$. For the NLO
anomalous dimensions we have used the simplified expression derived
by Yndurain et al. \cite{LopYnd79} which works with an accuracy
better then $0.2 \%$, namely
\begin{equation}
\gamma_j^{(1)} = \frac{1}{4} \sum_{\ell = 0}^{\infty}
\frac{ {\cal A}_\ell \ln (j + 1) + {\cal B}_\ell }{(j + 1)^\ell},
\end{equation}
with coefficients
\begin{eqnarray}
&&{\cal A}_0 = \frac{32}{27} ( 201 - 9 \pi^2 - 10 N_f ),
\
{\cal A}_1 = \frac{512}{9} ,
\
{\cal A}_2 = - \frac{256}{9} ,
\
{\cal A}_3 = \frac{1792}{27} ,
\
{\cal A}_4 = - \frac{256}{3} ,
\\
&&{\cal B}_0 = \frac{16}{9}
\left( - \frac{63}{4} - 134 \psi (1) + 6 \zeta (3)
- 7 \pi^2 + 6 \pi^2 \psi (1) \right)
+ \frac{32}{27} \left( \frac{3}{4} + \pi^2 + 10 \psi (1) \right) N_f,
\nonumber\\
&&{\cal B}_1 = \frac{16}{9} \left( 109 - 32 \psi (1)
- 3 \pi^2 - \frac{22}{3} N_f \right),
\
{\cal B}_2 = \frac{8}{9} \left( - \frac{1015}{3}
+ 32 \psi (1) + 7 \pi^2 + \frac{178}{9} N_f \right),
\nonumber\\
&&{\cal B}_3 = \frac{32}{27} \left( 263 - 56 \psi (1)
- \frac{9}{2} \pi^2 - 18 N_f \right),
\
{\cal B}_4 = - \frac{42692}{135} + \frac{256}{3} \psi (1)
+ \frac{236}{45} \pi^2 + \frac{1912}{81} N_f .\nonumber
\end{eqnarray}

In the following we will have a closer look on the distribution
functions evolved up to the reference scales $Q^2 = 10, 100\
{\rm GeV}^2$ for the skewedness parameters set equal to
$\zeta = 0.1, 0.5, 1.0$ and compare them with LO results.
We exploit for these purposes an expansion in terms of Legendre
polynomials. We perform the evolution\footnote{The calculations
were done with a code written for MAPLE.} by evaluating 70
moments in the series (\ref{PolExp}) in the case of the
Radyushkin's model distribution. In the particular case of the
asymptotic distribution the calculation was made in a different
way. Due to the support properties of the latter we used
the expansion of the non-forward distribution in Gegenbauer
polynomials $C_j^{3/2}(2x/\zeta -1)$, i.e.\ we directly employed
Eqs.\ (\ref{ConfExp},\ref{CorrConfWave},\ref{ConfWave}).
Thus there is no need for a double expansion which would
restrict the number of terms one can treat in the expansion. We
have used up to $100$ polynomials. In both cases we made a fit
to get smooth curves instead of rapidly oscillating ones.

The results are shown in Fig.\ \ref{Asydistribution},\ref{RadyushkinModel}.
As can be seen there is only a very small difference between NLO
and LO evolved distribution. As expected this deviation grows with
increasing $Q^2$, but it remains small even for large $Q^2$. Note
that for asymptotically large $Q^2$ the excitation of higher harmonics
will die out and both distributions take the form
${\cal O}_{\rm as} (x, \zeta)$ from Eq.\ (\ref{AsDis}).

%%%%%%%%%%%%%%%%%%%%%%%%%%%%%%%%%%%%%%%%%%%%%%%%%%%%%%%%%%%%%%%%%%%%%
%\section{Conclusion.}
%%%%%%%%%%%%%%%%%%%%%%%%%%%%%%%%%%%%%%%%%%%%%%%%%%%%%%%%%%%%%%%%%%%%%

\noindent {\it 4. Conclusion.} To conclude, in this note we presented
our results on the numerical evolution of the non-forward distribution
function in two-loop approximation in the flavour non-singlet
channel. We studied two models: the asymptotic function and Radyushkin's
model \cite{Rad98}, we have found that the net effects of NLO
corrections to the evolution kernels are extremely small and do not
exceed the level of a few percent for moderately large $Q^2$. These
results together with the fact that the evolution of the
${\cal O}_{\rm as} (x, \zeta)$ (\ref{AsDis}) is governed by
off-diagonal elements of the kernel in the basis of Gegenbauer
polynomials suggest that the latter are small as compared to diagonal
entries and, therefore, the kernel is quasi-diagonal in this basis.

\vspace{0.5cm}

One of us (A.B.) would like to thank A.V. Radyushkin for useful
correspondence and providing some of his results on the model of
the non-forward distributions before publication. He also what
to thank G.P. Korchemsky and B. Pire for useful conversations and
O.V. Teryaev for discussions about the positivity constraints for
non-forward distributions. A.B. is grateful to G.P. Korchemsky
and D. Schiff for the hospitality at LPTHE (Orsay) and B. Pire
at \'Ecole Polytechnique. A.B. was supported by the Alexander von
Humboldt Foundation and partially by Russian Foundation for
Fundamental Research, grant N 96-02-17631.

\end{document}